\documentclass[wssquare]{ws-procs961x669} 
\begin{document}
\title{Virial clouds evolution from the last scattering upto the formation of first stars}

\author{Noraiz Tahir$^{1, 2, \dagger}$,  Asghar Qadir$^{3, \dagger\dagger}$, Muhammad Sakhi$^{4,\dagger\dagger\dagger}$, and Francesco De Paolis$^{1, 2, \dagger\dagger\dagger \dagger}$}

\address{$^1$Department of Mathematics and Physics ``E. De Giorgi'', University of Salento, Via per Arnesano, I-73100 Lecce, Italy\\
$^2$INFN, Sezione di Lecce, Via per Arnesano, I-73100 Lecce, Italy \\
$^3$Abdus Salam School of Mathematics, G.C. University, Lahore, Pakistan \\
$^4$Department of Physics, Quaid-e-Azam University, Islamabad, Pakistan\\
$^\dagger$noraiz.tahir@le.infn.it\\
$^{\dagger\dagger}$asgharqadir46@gmail.com\\
$^{\dagger\dagger\dagger}$sakhi.cosmos@gmail.com\\
$^{\dagger\dagger\dagger\dagger}$ francesco.depaolis@le.infn.it}

\begin{abstract}
The asymmetry in the cosmic microwave background (CMB) towards several nearby galaxies detected by {\it Planck} data is probably due to the rotation of ``cold gas'' clouds present in the galactic halos. In 1995 it had been proposed that galactic halos are populated by pure molecular hydrogen clouds which are in equillibrium with the CMB. More recently, it was shown that the equillibrium could be stable. Nevertheless, the cloud chemical composition is still a matter to be studied. To investigate this issue we need to trace the evolution of these virial cloud from the time of their formation to the present, and to confront the model with the observational data. The present paper is a short summary of a paper \cite{noraiz2021epjc}. Here we only concentrate on the evolution of these clouds from the last scattering surface (LSS) {\it up to} the formation of first generation of stars (population-III stars). 
\end{abstract}


\bodymatter

\section{Introduction}
The study of the nature of galactic halos and their dynamics is a task that is difficult to address. Here we present a summary of a more detailed analysis addressing this issue \cite{noraiz2021epjc}. 

The $\Lambda$ cold dark matter ($\Lambda$CDM) model entails that  $\sim 5\%$ of our Universe is made of baryonic matter \citep{ade2016planck}, of which $\sim 60\%$ is detected \citep{kassin2006dark, nicastro2017decade, dave2001baryons, li2018baryon, cen1999baryons}, but $\sim 40\%$ is still undetected at present. This is the so-called ``missing baryon problem''.

In 1995 it was proposed that a fraction, $f$, of these missing baryons is present in the galactic halos in the form of {\it pure} molecular hydrogen ($H_2$) clouds which are in equillibrium with the CMB \citep{de1995case}. The difficutly was on the observation of such ``chameleons'' megerd with the background. One of the suggestions was, to look for a Doppler shift effect due to the rotation of galaxies, assuming that the rotation of these clouds is synchronized with the rotation of galactic halos, and hence they should be Doppler shifted, those clouds roatating towards us should give a blue-shift effect while those rotating away from us would give a red-shifted contribution.

In 2011 WMAP data was analyzed for M31. The analysis revelaed a temperature asymmetry in the CMB which was almost frequency independent \citep{de2011possible}, which was a strong indication of the Doppler shift effect due to the galactic halo rotation. This opens up a window to observe these clouds and to study the baryonic content of the galactic halos \citep{de2012cmb}. Soon after WMAP, in 2014 {\it Planck} data towards M31 was analyzed and  the asymmetry was seen at a more precise level \citep{de2014planck}. A temperature asymmetry was also detected towards several nearby spiral galaxies \citep{de2015planck, gurzadyan2015planck, de2016triangulum, gurzadyan2018messier}. 

There was more than one item of evidence of the predicted Doppler shift effect, but observing that there is a Doppler shift due to the halo rotation does not reveal the true nature of the effect, i.e. if it is partially or fully due to the molecular clouds in the halos, or, if is there anything else that could give a masking or mimiking effect in the asymmetry. Another unanswered question is the chemical composition of these clouds: are they pure $H_2$ clouds, or there is some contamination of dust or heavier molecules in them? It is quite obvious that galactic halos contain a significant fraction of dust that {\it should} contaminate these clouds \citep{yershov2020distant}, so one needs to model the clouds. 

As these clouds should survive on acount of the virial theorem, they were called ``virial clouds''. These clouds were modeled and it was seen that at the current CMB temperature the centeral density of pure $H_2$ clouds was $\approx 1.60 \times 10^{-18} {\rm kg~ m^{-3}}$, similarly the mass and radius were $\approx 1.93 \times 10^{-4} M_{\odot}$, and $\approx 0.032$ pc \citep{qadir2019virial}. The change in their physical parameters with the contamination of heavier molecules and dust were also estimated. It was seen that as the contamination of dust and hevier molecules were increased in the clouds, they became denser, and their mass and radius decreased \citep{qadir2019virial, tahir2019seeing, tahir2019constraining}. An objection was raised on the stability of these clouds, as it was believed that molecular clouds can not be stable at this, very low, CMB temperature, as there would be no mode that could be excited by the photons and the cloud might collapse to form stars or other planetary objects \citep{padmanabhan1990statistical}, but it was demonstrated that this equilibrium {\it does} arise on account of the
translational mode, despite its extremely small probability, because of the
size of the virial clouds and the time scales available for thermal equilibrium to be reached, so that the time required for thermalization is much less than that required for collapse \citep{qadir2019virial}.

Modeling the virial clouds and estimating the change in physical parameters with the contamination of heavier molecules and dust, and observing the CMB temperature asymmetry by {\it Planck} data still does not answer the question on the nature of virial clouds and the exact cause of the observed asymmetry in the CMB. One has to run the clock back and trace the evolution of virial clouds when they were formed at the LSS to the present. This task needs to be done in two phases: $(i)$ from LSS up to the formation of population-III stars; and $(ii)$ from the formation and explosion of population-III stars to the present. Here we discuss the first part of evolution, as various qualitative changes took place during the formation of population-III stars \cite{jeon2014recovery}. Hence there will be significant changes during the second step, which will be studied more clearly and in more detail later.

\section{First epoch of virial clouds evolution \label{epoch}}
Virial clouds would have formed at $z=1100$, the last scattering time (LSS) and they evolved in their chemical composition and physical parameters,  but in order to maintain their stability they survived the collapse and stayed in quasi-static equilibrium with the CMB since then. It is quite obvious that when they formed they should have had the primordial chemical composition, i.e. $\sim 75\%$ atomic hydrogen and $\sim 25\%$ helium. In addition to H and He there were other atoms and molecules like deuterium, helium-3, lithium and molecular hydrogen, which could have contributed to the virial clouds, but their fraction was negligilble as compared to H and He. Hence these molecules and atoms could not have any significant effect on the virial cloud physical parameters. As a result the ratio of atomic hydrogen and helium would remain the same and these would be the main component to form the virial clouds during the period, but there should be a fast change in their chemical composition after the formation and explosion of population-III stars. 
\subsection{Hydrogen-Helium virial clouds \label{HHe}}
Since virial clouds must be considered to be in thermal equilibrium because they are embedded in the heat bath of the CMB, we need to use the
canonical distribution function for a fixed temperature and use the cooling
of the heat bath to provide a quasi-equilibrium. Moreover, these clouds should start to form in the potential well of cold dark matter (CDM). As the clouds are thermalized the potential well will not cause them to collapse to form population-III stars \cite{barkana2001}, but will modify the physical parameters of virial clouds.

To obtain a general expressions we consider a virial cloud composed of an arbitrary mixture of H and He, with mass fractions $\alpha$ and $\beta$. Then, we use the primordial cosmological  fractions of H and He for the final computation. The total mass of the cloud is, obviously,  $M_{cl}(r)=\alpha M_{H}(r)+\beta M_{He}(r)$,  with the condition $\alpha+\beta=1$. The density distribution for two fluids is given by \citep{qadir2019virial}

\begin{align}
	&\rho_{cl}(r)=\sqrt{\frac{64}{27}}\frac{(G\rho_{c_H}\rho_{c_{He}})^{3/2}}{(k_BT)^{9/2}}(m_Hm_{He})^{5/2}\nonumber\\
	&exp\left[-\frac{1}{2}\left(\frac{\alpha G M_H(r)m_H}{rk_BT}+\frac{\beta G M_{He}(r)m_{He}}{rk_BT}\right)\right].
	\label{twofluiddistributionequation}
\end{align}

and the corresponding differential equation can be written as \citep{qadir2019virial}
\begin{align}
	r\frac{d\rho_{cl}(r)}{dr}&-r^2\left(\frac{2\pi G}{k_BT}\right)[\rho_{cl}(r)(\alpha \rho_{c_H}m_H\nonumber \\ &+\beta \rho_{c_{He}}m_{He})]-\rho_{cl}(r)\ln\left(\frac{\rho_{cl}(r)}{\tau}\right)=0,
	\label{twofluiddifferentialequation}
\end{align}
where, $\tau=(8/3\sqrt{3})[(G\rho_{c_H}\rho_{c_{He}})^{3/2}/(k_BT)^{9/2}][m_Hm_{He}]^{5/2}$, $\rho_{c_{H}}$ is the central density of H cloud, $\rho_{c_{He}}$ is the central density of He cloud, $m_H$, the mass of single atom of hydrogen, and $m_{He}$, the mass of single atom of helium.

We use eq.(\ref{twofluiddifferentialequation}) to estimate the central density of the clouds. We estimated the central density, Jeans mass and radius of the two-fluid virial clouds with primordial fraction of H, and He, i.e. $\alpha= 0.75$, and $\beta=0.25$. In order to solve eq.(\ref{twofluiddistributionequation}) numerically, we assumed a guess value of $\rho_{c}$, at a fixed temperature, and estimate where the density becomes {\it exactly} zero at the boundary. We then compare the Jeans radius with that central density with the value available to us. We adjust the central density so that the density becomes zero exactly at the Jeans radius. In this way we get a self-consistent solution of the differential equation subject to the given boundary conditions. Next, we decrease the temperature and repeat the process. 

It was seen that with the decrease in temperature and redshift, the density of virial clouds increased, at $z= 1100$ these clouds were $\approx 8.53 \times 10^{-20} kgm^{-3}$ and at $z=50$ the density of the clouds was $\approx 6.05 \times 10^{-19} {\rm kg~m^{-3}}$. On the other hand the mass decreased with the decrease in temperature and so as the radius of the clouds. These clouds were massive $\approx 50$ pc, with a mass $\approx 3.05 \times 10^5 M_{\odot}$ at $z=1100$, then at $z=50$ the clouds were $\approx 3.54$ pc with a mass of $\approx 7.54 \times 10^3 M_{\odot}$. 

\section{Results and discussion \label{summary}}
The first stage of the virial cloud evolution turns out to be quite simple. We have seen that virial clouds became denser with time, and they lost mass and shrunk in size. The physics of the second stage of evolution will not be that simple since there will be cooling due to the formation of molecular hydrogen, turbulence and the angular momentum effects in the virial cloud after the formation and explosion of population-III stars. We need to consider all these things while tracing the second epoch of evolution, and this will be done in the later paper.

One could expect molecular hydrogen to be formed in the first epoch, but during the era of study in the current summarized paper any $H_2$ molecules formed in the clouds will be unstable and dissociate due to the radiation pressure at that time, since $H_2$ needs dust particles to remain stable. The molecular cooling of such clouds played a vital role after the recombination and  thir cooling rate has been analyzed from $T\sim 120- 3$ K  \cite{1999A&A...345..723P}. Hence, we do not need to consider the effect of cooling from molecular hydrogen till population-III stars exploded. It may also be expected that the electronic transition mode in hydrogen and helium atoms by the CMB photons. One can check that most of the CMB photons at 3000 K do not have enough energy to excite the electrons of a hydrogen atom from its ground state, but the higher energy tail does have sufficient energy to do so. The percentage of such photons is $0.016$, which is quite low. Hence, there will be a negligible effect of this mode, but these effects will be more significant in the later stage of evolution.

\section*{Acknowledgements}
	FDP and NT acknowledge support from the TAsP and Euclid INFN projects.

\end{document}